\documentclass[%
 reprint,
 amsmath,amssymb,
 aps,
]{revtex4-1}

\usepackage{subfigure}
\usepackage{color}
\usepackage{graphicx}
\usepackage{dcolumn}
\usepackage{bm}
\usepackage{hyperref}

\newcommand{\iso}[2]{{\ensuremath{{}^{#2}\mathrm{#1}}}}

\begin{document}

\preprint{PRC}

\title{High precision half-life measurement of \iso{Sm}{147} $\alpha$ decay from thin-film sources}

\author{H. Wilsenach}
 \email{Corresponding author: \\ heinrich.wilsenach@tu-dresden.de}
\author{K. Zuber}
\affiliation{%
Institut f\"ur Kern- und Teilchenphysik, \\ Technische  Universit\"at Dresden,  01069 Dresden, Germany
}%

\author{D. Degering}
\affiliation{%
VKTA - Radiation Protection, Analytics \& Disposal Rossendorf e.V.,\\ P.O.Box 510119
}

\author{R. Heller}
\affiliation{%
Institute of Ion Beam Physics and Materials Research, \\ Helmholtz-Zentrum Dresden-Rossendorf ,\\ Bautzner Landstr. 400, 01328 Dresden 
}%

\author{V. Neu}
\affiliation{%
Institute for Metallic Materials, \\ IFW Dresden, Helmholtzstrasse 20, 01069 Dresden, Germany
}%



\date{\today}

\begin{abstract}

An investigation of the $\alpha$-decay of \iso{Sm}{147} was performed using an ultra low-background Twin Frisch-Grid Ionisation Chamber (TF-GIC). Four natural samarium samples were produced using pulsed laser deposition in ultra high vacuum. The abundance of the \iso{Sm}{147} isotope was measured using inductively coupled plasma mass spectrometry.  A combined half-life value for \iso{Sm}{147} of $1.079(26) \times 10^{11}$~years  was measured. A search for the $\alpha$-decay into the first excited state of \iso{Nd}{143} has been performed using $\gamma$-spectroscopy, resulting in a lower half-life limit of $T_{1/2} > 3.1 \times 10^{18}$ years (at 90\% C.L.).

\begin{description}
\item[Usage]
Secondary publications and information retrieval purposes.
\item[Key words]
\iso{Sm}{147},
Samarium,
Half-life,
Frisch-Grid Ionisation Chamber,
Geochronology,
Dating.
\item[PACS numbers]
23.60.+e, 29.40.Cs, *91.80.Hj, 
\end{description}
\end{abstract}

\pacs{Valid PACS appear here}
\keywords{Suggested keywords}
\maketitle


\section{\label{sec:intro}Introduction}

Various long living nuclides are used for dating of the early solar system and the universe. Due to the relatively long half-life of \iso{Sm}{147} which is of the order of the age of the Universe, the isotope is commonly used as a cosmo-chronometer. Here, samples are dated using the relative abundance of \iso{Nd}{143} to \iso{Nd}{144} \cite{IPA}, with the first isotope stemming from the $\alpha$-decay of \iso{Sm}{147}. The
relative ratio increases as a function of time due to the shorter half-life of \iso{Sm}{147}. This dating method is vulnerable to the measured half-life of \iso{Sm}{147}. Any discrepancy and uncertainty in the half-life values should therefore be investigated thoroughly.

This paper describes a high precision investigation of the isotope \iso{Sm}{147} ($J^{\pi}=7/2^-$) $\alpha$-decay into the ground  and first excited state of \iso{Nd}{143}  using $\alpha$-spectrometry and $\gamma$-spectroscopy. The isotope \iso{Sm}{147} is a well known calibration source for $\alpha$-spectrometry with a Q-value of  2.3112(10)~MeV~\cite{ame2012}, resulting in an $\alpha$-peak at 2.248(1)~MeV. In addition, \iso{Sm}{147} has a reasonably well measured half-life with a weighted mean of $1.060(11) \times 10^{11}$~years \cite{A147} for the ground state transition but no experimental limits exist for excited states, see half-life compilation in \cite{kos09}. The latest measurement on its half-life was performed with liquid scintillators, where the quenching factor has to be known. 

The purpose of this paper is a high precision half-life measurement using a low-background ionisation chamber and thin-films as samples to explore the claimed precision.

\section{Sample Production and Characterisation}

One of the primary obstacles in the measurement of $\alpha$-decays from a solid target is high self absorption. This is due to the $\alpha$-particles low energy to charge mass ratio. The $\alpha$-particles quickly lose most of their energy within the sample through ionisation. The $\alpha$-particles from natural decay have an energy of less than 10~MeV, the maximum range of an $\alpha$-particle with this energy is of the order of $\mu$m for most solid materials. This fact makes it difficult to measure the $\alpha$-decay using thick targets, as most of the $\alpha$-particles will not be detected, or the energy will be smeared over a wide range resulting in a low-energy tail in the spectrum. To overcome this major limitation the targets used in this work are made as thin as possible, while still having enough material to ensure a measurable activity with good statistics. 

The technique used in the sample preparation for this work is laser deposition under an ultra-high vacuum. This method is described in \cite{neu}. Other than working in a background gas pressure as e.g. during magnetron sputtering, the ulta-high vacuum ensures a ultra-clean environment which  avoids contaminations of the films. An off-centered substrate rotation during film growth leads to a very homogeneous deposition ($<$ 1\% thickness variation). The current drawback is that only a small sample area can be produced, but this limitation can be overcome in the future. Four 1$\times$1~cm$^2$ natural samarium samples were produced with this method. The following thicknesses were achieved; 34.6(13) ~nm, 41.44(61)~nm, 232.7(64)~nm and 911.8(93)~nm. Table~\ref{tab:sample} gives more information on the sample thicknesses. 

The sample geometry was characterised as follows. The number of atoms per area was obtained using the Rutherford back-scattering (RBS) method \cite{neu}. This method uses an incident beam of $\alpha$-particles to probe the samples. The ion beam has a small probability of interacting with the Coulomb potential of the atomic core and thus being scattered. The $\alpha$-particle is scattered at a chosen angle with a specific energy. Analysis of this spectrum gives information of the layer areal density and to some extend also on the structure of the samples. With the assumed density of 7.54~g/cm$^3$\cite{Density} for samarium, the thickness can be derived. For further reading on RBS see \cite{oura2013surface}. The samples are also scanned on multiple positions to determine a surface homogeneity. Additionally energy-dispersive X-ray spectroscopy (EDX) was used to determine the layer's composition and thickness. This was done to validate the RBS measurement, but it cannot be used to determine the number of atoms with the same precision as RBS.

The RBS measurement only reveals the number of atoms per unit area. The total number of samarium atoms can only be delivered by multiplying the RBS value with the area of each sample. The total area of the sample was therefor measured by the use of a VHX-2000 Digital Microscope. The microscope takes an ultra-high resolution image of the deposition surface. This image is then analysed with digital software to determine the deposition area.

The surface was also scanned with a Profilometer to validate the area measurements and to give some insight into the homogeneity of the coating. The thickness profile was scanned at several paths for each sample, giving the roughness of the sample.

\begin{figure*}[t]
\setlength{\abovecaptionskip}{17.5 pt plus 4pt minus 2pt}
\mbox{
\subfigure{\includegraphics[width=0.35\linewidth,height=6cm]{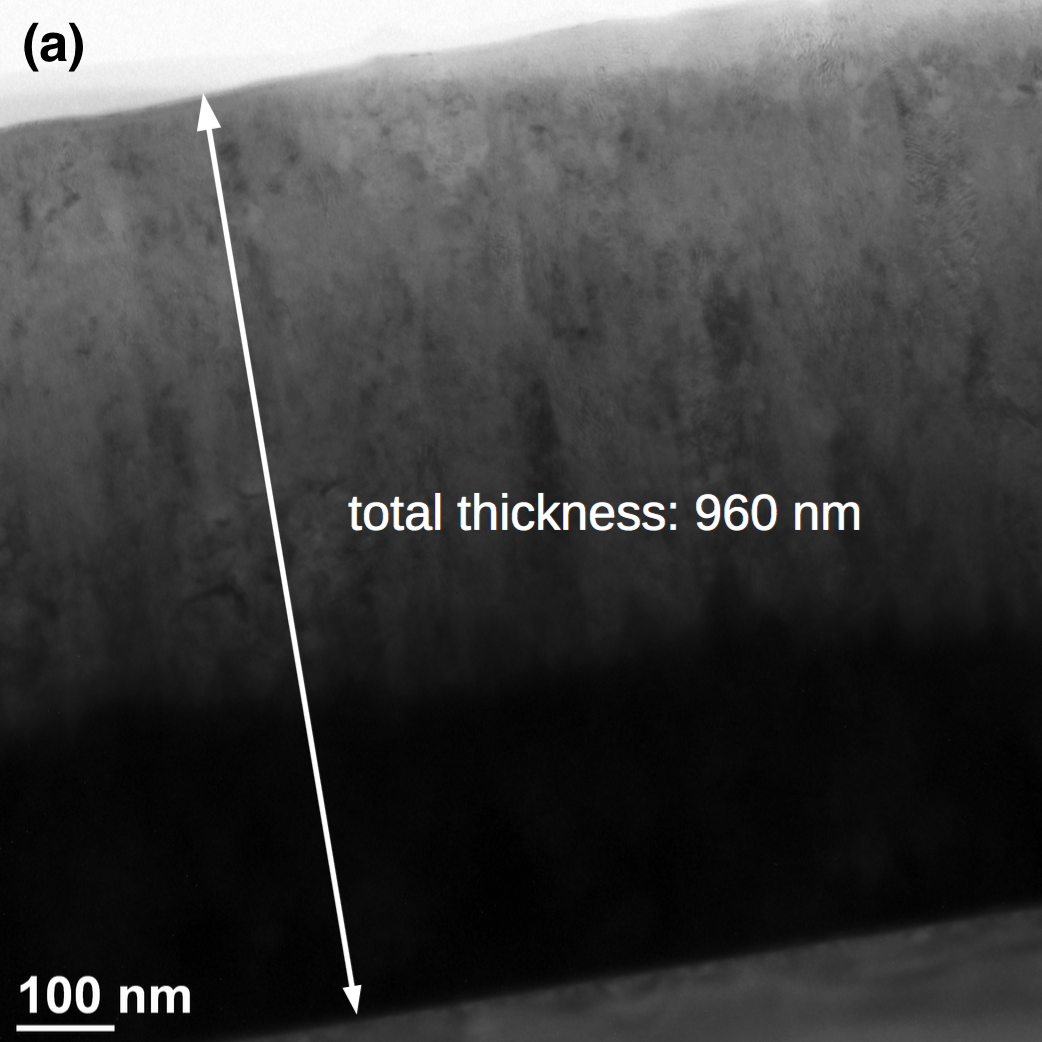}}\,
~\hspace*{1cm}~
\subfigure{\includegraphics[width=0.35\linewidth,height=6cm]{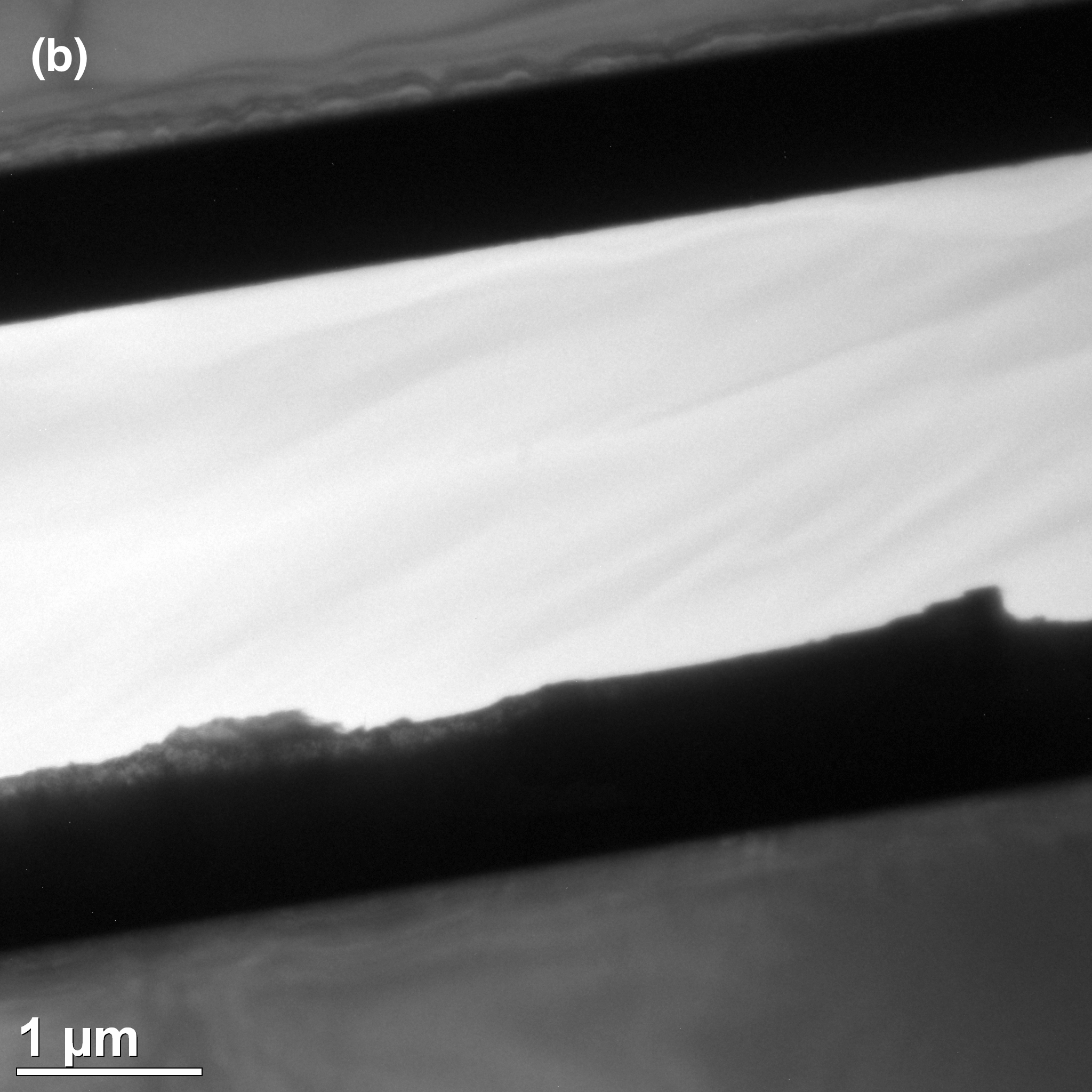}}
}
\vspace{-0.6 cm}
\caption{Transmission electron microscopy images of sample SM004. Image (a) shows the measured thickness which is 938.5~nm with the chromium layers subtracted. The layer is even and homogeneous. Image (b) shows the comparison of different pieces of the same sample, here shown as the two horizontal dark lines. The white in the centre of the image is the glue used in the TEM preparation. Here it is easy to see that some structure is visible on some part of the surface. This validates the increased roughness observed in the RBS investigations. }
\label{fig:TEM}
\end{figure*}

The thicker samples showed an unusually high roughness in the RBS spectra. To verify this roughness transmission electron microscopy (TEM) was performed on sample SM004. In this characterisation a thin lamella of the sample is cut out as a cross section and successively investigated using a TEM. This method gives an insight into the sub structures formed during the coating. Figure~\ref{fig:TEM} shows the corresponding TEM images of sample SM004 at two different magnifications. The overall structure is smooth, and it is possible to see the exact thickness of the samarium coating as well is the chromium used for the adhesion layer on the bottom and the protection layer on the top. The TEM showed some thickness variations on top of the coating which was also measured with the RBS. This validates the increased roughness observed in the RBS investigations. The thickness obtained from the TEM investigation is 938.5~nm. The measurement also reveal that no substructures were formed during the coating process, and that the layer can be treated as a solid film. Table~\ref{tab:sample} contains the measured values of the sample characterisation for all four samples. The density measured for SM004 from the combination of RBS and the thickness of the TEM is 7.59(7)~g/cm$^3$. This value is in good agreement with the assumed density of 7.54~g/cm$^3$. 

\begin{table}[htbp]
\caption{Table containing the results of the characterisation of the samples. The uncertainty on the EDX is assumed to be 5\%, the value is only used for comparison. The thickness values here are derived with the assumed mass density of 7.54~g/cm$^3$\cite{Density}.}
\begin{tabular}{|l|c|c|c|c|clc|}
\hline
Sample  & Area [cm$^2$] & Surface density  & \multicolumn{2}{c|}{Thickness} \\
 Name&  & [$10^{15}$/cm$^2$] &  EDX [nm] &  RBS [nm]\\ \hline
SM001 & 0.8027(4) & 106(2) & 31.4(16) & 34.6(13)  \\ \hline  
SM002 & 0.8014(1) & 123	(3) & 40.2(20) &  41.44(61)  \\ \hline
SM003 & 0.7763(10) & 680(7)& 232(12) & 232.7(64)  \\ \hline 
SM004 & 0.8047(2) & 2720(22)& 822(41) & 911.8(93)   \\ \hline  
\end{tabular}
\label{tab:sample}
\end{table}

Sample SM003 was prepared with a stripe of carbon placed on top of the substrate prior to deposition, so that samarium and chromium could be locally lifted-off afterwards by washing the sample in acetone. The step height was measured by atomic force microscopy (AFM), which resulted in a samarium layer thickness of 241(8) nm. This value is in good agreement with the EDX and RBS values.

The techniques described above used to characterise the samples are well established methods for the investigation of thin films. They also have the added advantage of being complimentary to each other. This is important for the measurement as it means that the obtained value does not solely rely on one unvalidated method, which could introduce an unknown systematic uncertainty into the result. 

Natural samarium consists of seven isotopes. The isotope of interest \iso{Sm}{147} has a literature natural isotopic abundance of 14.99(18)\% \cite{deLaeter2003}. To measure the real isotopic abundances of the samples, inductively coupled plasma mass spectrometry (ICP-MS) was performed on a small part of the material used in the production of the samples. This measurement was done at Atomki Debrecen. The  isotopic abundance was measured and is shown in Tab.~\ref{tab:icpms}. The value measured for \iso{Sm}{147} is in agreement with the natural abundance tabulated in \cite{deLaeter2003}.

\begin{table}[htbp]
\caption{Table of measured sample isotopic abundances. The measurement was done with ICP-MS. The values are based on the average and standard deviation of 10 separate measurements. (Literature values taken from \cite{deLaeter2003}.)}
\begin{tabular}{|c|c|c|}
\hline
Isotope & Measured Abundance  & Literature Abundance \\
 & [\%]  & [\%]  \\ \hline
\iso{Sm}{144}  & 3.01(1)  &  3.07(7) \\ \hline
\iso{Sm}{147}  & 14.77(5) & 14.99(18) \\ \hline
\iso{Sm}{148}  & 11.22(4) & 11.24(10) \\ \hline
\iso{Sm}{149}  & 13.91(5) & 13.82(7) \\ \hline
\iso{Sm}{150}  & 7.08(3)  &   7.38(1) \\ \hline
\iso{Sm}{152}  & 26.47(9) &  26.75(16) \\ \hline
\iso{Sm}{154}  & 23.5(1)  &   22.75(29) \\ \hline
\end{tabular}
\label{tab:icpms}
\end{table}

\section{Experimental Setup}

To accurately measure an activity in the order of 100~counts~per~day (c.p.d) an ultra low-background Twin Frisch-Grid Ionisation Chamber (TF-GIC) was used \cite{cham}. The chamber is constructed from radio pure materials to lower the intrinsic radioactive background. The design is based on two mirrored detectors sharing an anode where the upper chamber is used as a veto system against noise and cosmic rays. Samples are placed on the bottom of the lower chamber. The gas used in the chamber is P10 (90\% Ar, 10\% CH$_{4}$). The data acquisition (DAQ) system uses a CAEN~A1422 Low Noise Fast Rise Time Charge Sensitive Preamplifier coupled to a CAEN Mod. N6724 4 Channel 14bit 100MS/s fast analog to digital converter (FADC). An FADC samples the pulse shape at 10~ns intervals for 2~$\mu$s. For more information on the set up and characterisation of the TF-GIC see \cite{cham}, the calibration of the chamber is described there in more detail as well.


In addition, pulse shape discrimination, described later in more detail, is used to separate signal from background events, the efficiency of which
is 99.3\%. The overall detection efficiency of the chamber has been measured with an \iso{Am}{241} source. The total detection efficiency was found to be 98.6(22)\%, the efficiency does not include the loss of signal due to the geometry of the sample. Additionally, a set of cuts were developed based on the physical properties of the signal carriers in the ionisation medium. Combining the chambers low internal contamination with the high false signal discrimination gave an overall background rate of 10.9(6)~c.p.d in the energy region of 1~MeV to 9~MeV. The majority of this background is from a tiny \iso{Rn}{222} contamination, which is above 4~MeV. During a 30.8~day background run only 16 events were measured inside the region of 1~MeV to 3~MeV.\\

\begin{figure}[htbp]
	\begin{center}
		\includegraphics[width=8.6cm]{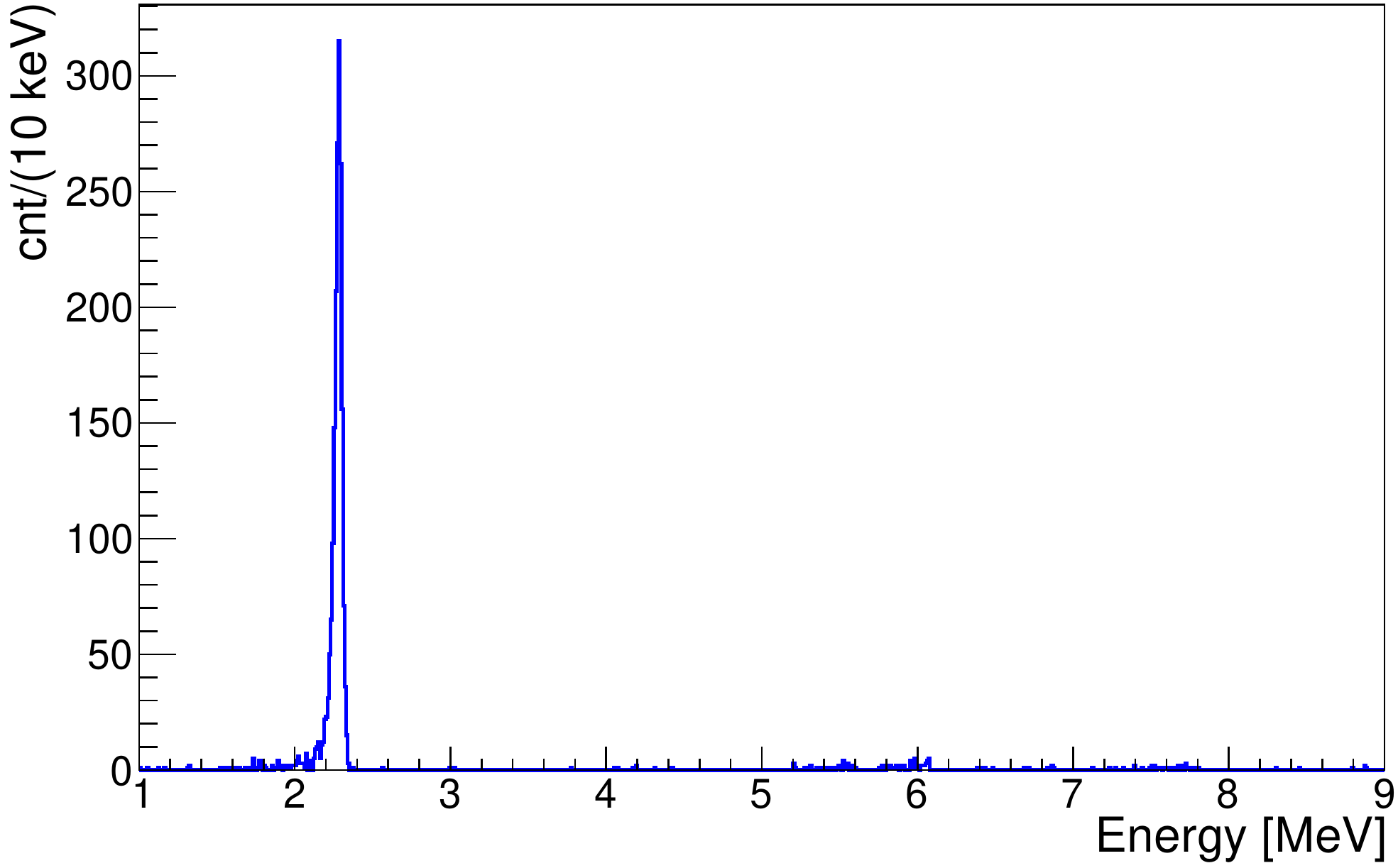}
		\caption{Energy spectrum measured  for 19~days of the 34.6(13)~nm thick sample, including background after quality cuts. The measured $\iso{Sm}{147}$ peak shows the expected maximum at 2.28~MeV. All events above 2.5~MeV are from Th and U contaminations, the majority of which are above 5~MeV. The energy resolution of the peak is $\sigma_{E}$ = 16.76(88)~keV.}
		\label{fig:needle}
	\end{center}
\end{figure}

The measurement was performed with the four samples mentioned before. Each sample was measured multiple times. The measured peak of a typical run is shown as an example in Fig.~\ref{fig:needle}.

\section{Analysis}

Before extracting the half-life, various aspects have to be considered. To determine the effect of systematic uncertainties on the measurement, four samples were produced. The samples were made to cover a range of thicknesses. Each sample was produced using a similar method. The only parameter that was varied was the thickness of the samarium coating. Each sample is then characterised using the techniques mentioned above. The decay rates were then measured in the TF-GIC.

The first source of systematic uncertainty is in the correct determination of the number of atoms in the sample. This has a direct effect on the value of the half-life, and is one of the main objectives to understand for the measurement. For this, characterisation methods were chosen that could compliment each other. If the thicknesses as well as the estimated  statistical uncertainty of the thicknesses were determined correctly, then the half-life values  for all four samples should be the same. 

The second cross check for the validity of the result is the rate expected for each sample. If the characterisation of each sample is successful, then each sample should have a relation between the measured rate and the number of \iso{Sm}{147} atoms. To determine this, a half-life is calculated per sample. These values are compared. If there is an unexpected systematic error, there will be a discrepancy in the half-life values for each sample.

Another source of systematic uncertainty is the tailing caused by the self ionisation of the $\alpha$-decay. To take this into account a simulation was created using the GEANT4 simulation package (version 10.2.0) \cite{GEANT4}. The energy tail was used as a free parameter in the fitting. This was done to give an extra cross check for the RBS, AFM, EDX and TEM. This was important, because the GEANT4 simulations also give the detection efficiency which is a function of the geometry of the sample. Deriving the thickness from the energy tailing is extremely sensitive to the energy smearing and calibration of the spectrum. For this reason every spectrum was re-calibrated to match the energy of the Monte Carlo spectrum. 

The samples were measured on two holders. One was made of an radio-pure plastic called delrin. The other was composed of silicon. Both of the holders had very little background, but the silicon was more pure, due to it being a single 30~cm diameter wafer. 

\subsection*{Cuts}

The data analysis is based on pulse shape discrimination. The pulse shape of each event is analysed and stored in ROOT files \cite{root}. The basic parameters, pulse heights and trigger times are then extracted for each event. From these basic parameters it is possible to determine the energy deposited in the gas, the position of the centre of charge and the velocity of the ionised electrons in the gas. Cuts that are sensitive to these quantities have been developed and are applied to the data. The cuts ensure that only $\alpha$ like events that come from the sample holder are accepted. There are some events from surface contamination, but their energy is much higher than the $\alpha$-energy of \iso{Sm}{147}. Furthermore the background in the region of interest (ROI) of the signal is expected to be much smaller than the signal. The ROI is defined as $1\;\text{MeV} < E < E_{\alpha}+3\cdot\sigma$, where $E_{\alpha}$ is the measured energy of the $\alpha$-decay, and $\sigma$ is the energy resolution at $E_{\alpha}$.

The first cut applied to the data is the chamber selection cut. This cut discriminates the events from the top and bottom chamber by requiring no sizeable signal from the top chamber. In addition any noise will produce a signal on all of the FADC channels, therefore this cut is applied first as it has the largest noise rejection efficiency. The next cut selects the detector region of the event in the lower detector. This cut uses the fact that the event will occur in the interaction region (cathode to grid) before the detection region (grid to anode). Therefore the grid signal should trigger the FADC before the anode signal. These two cuts are the most basic cuts and their purpose is to remove as much noise as possible.

\begin{figure}[htbp]
\begin{center}
	\includegraphics[width=8.6cm]{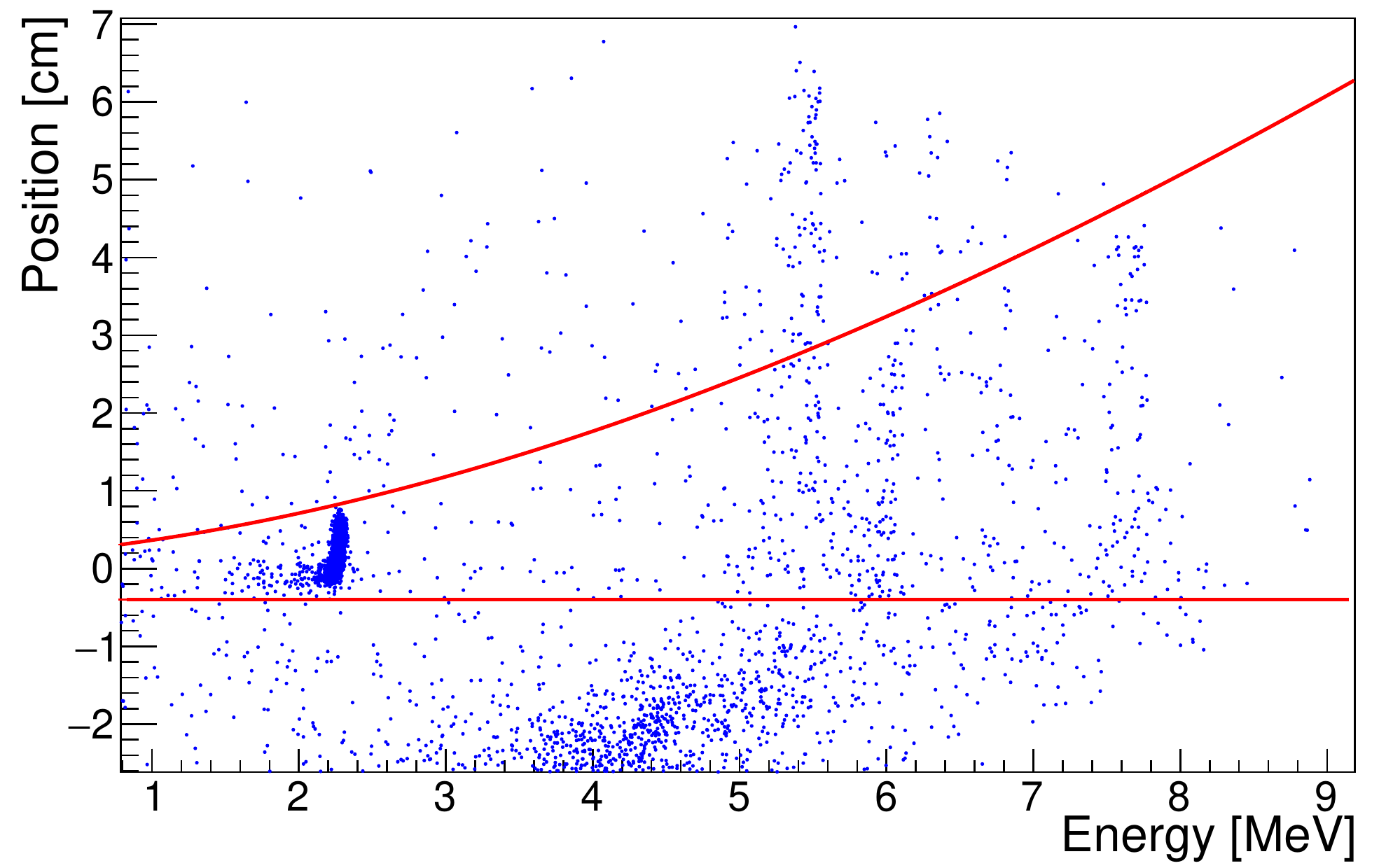}
\caption{Figure showing the position of each event in a typical run as a function of energy. The solid lines are the maximum and minimum of each cut. The events that fall between the solid lines are accepted. If an event occurs on the wall of the lower chamber its position is reconstructed above the maximum position. This can be clearly seen in the line from \iso{Rn}{222} at around 5.5~MeV. The events that have a position lower than the minimum around 4~MeV are from decays on the walls close to the anode. The events close to the maximum of the position cut have a trajectory at a tangent to the surface of the sample. The events close to the minimum of the cut are almost parallel to the surface of the sample. }
\label{fig:pos}
\end{center}
\end{figure}

The next set of cuts is based on the physical properties of the detection method. The first of which is the cut on the position of the event. Due to the design of the TF-GIC, every pulse shape carries position information. A simulation has been performed to determine the maximum range of an $\alpha$-particle of a certain energy inside P10 gas using SRIM \cite{SRIM}. This was cross-checked with GEANT4. If it violates the energy-range relation, it is not accepted as signal. This is shown in Figure~\ref{fig:pos}.

\begin{figure}[htbp]
\begin{center}
	\includegraphics[width=8.6cm]{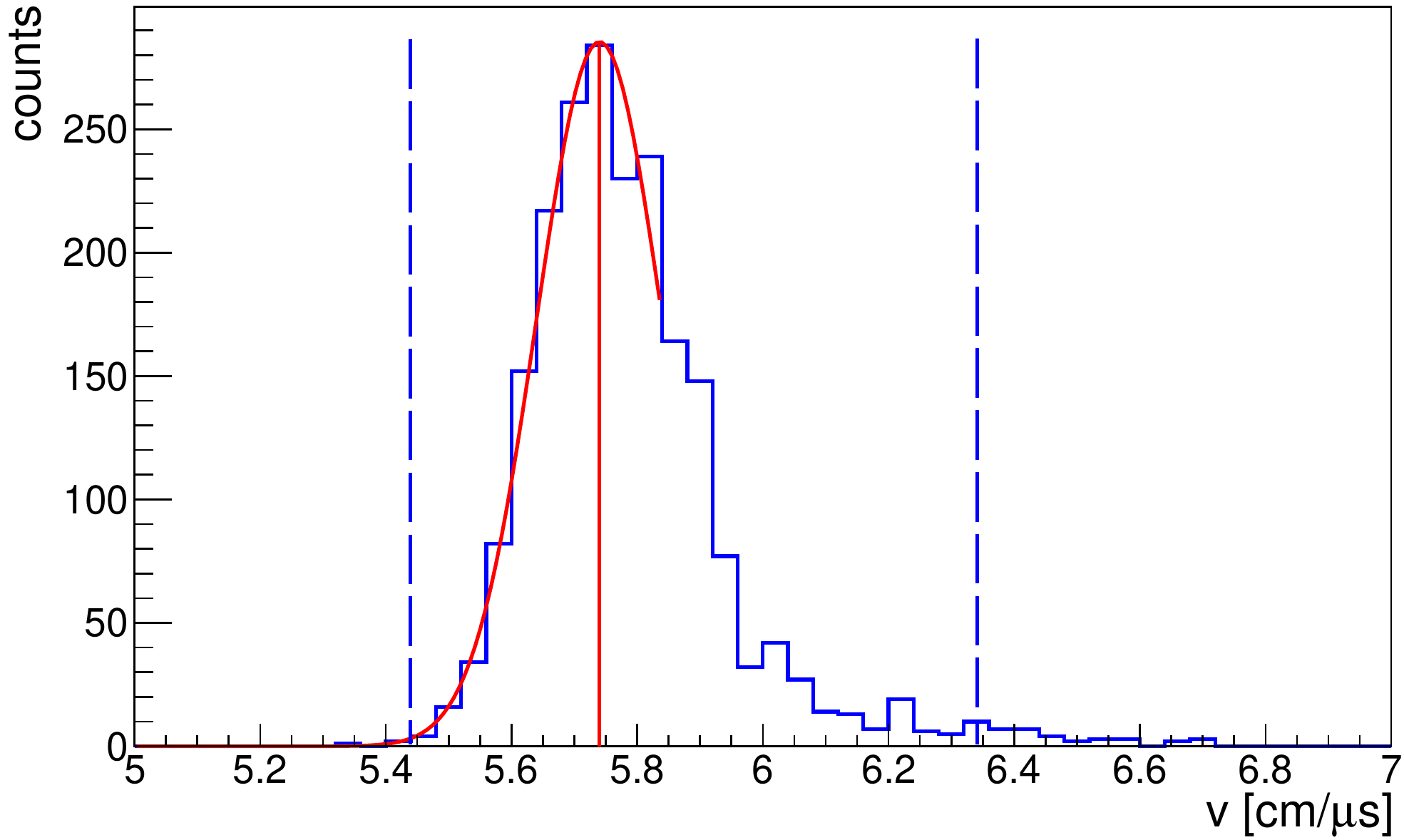}
\caption{Distribution of electron drift velocities $v$ in cm per $\mu$s. The values have been corrected to remove dependence on energy oand runtime, the position cut has also been applied. The solid red vertical line is the centre of the Gaussian fit is called the average drift velocity ($X_0$). The dashed blue vertical lines are the minimum ($X_0-3\cdot\sigma$) and maximum ($X_0+6\cdot\sigma$) of the drift velocity cut. Due to slight variations in the gas composition for each run, $X_0$ and $\sigma$ are taken from the fit.}
\label{fig:mu}
\end{center}
\end{figure}

The final cut is on the drift velocity $v$ of the signal carriers in the gas. This is determined using the position measurement of each pulse and the triggering information. This value is derived in such a way that there is no energy dependence in this cut. The drift velocity decreases slightly as a function of runtime (1.4\% drop per day) due to a small amount of outgassing in the inside of the chamber. This it not thought to be caused by an oxygen leak into the chamber as the energy is stable as a function of time. Moreover the chamber is run in overpressure. This can be described by an inverse first order polynomial and is corrected for in the DAQ. The velocity distribution has a Gaussian shape with a tail towards higher velocities shown in Figure~\ref{fig:mu}. The lower and higher value of the cut is set in units of sigma of the fitted Gaussian. The lower cut value is three sigmas smaller than the average drift velocity, where as the higher cut value is six sigmas higher than the average drift velocity. The number of sigmas used in each cut was calibrated using data from an \iso{Am}{241} source. Due to slight variations of the gas composition for each run, this has to be determined on a run by run basis. 

For every run the ionisation energy is constant as a function of time. This would not be the case if there was a problem with the preparation of the run, e.g. the flushing malfunctioned and some air entered the detector and mixed with the ionisation gas. The energy as a function of time is fitted using the least square method with a second order polynomial. The uncertainty on the energy is taken as the standard deviation of the anode pulse height sampled over a range of 1000 points. Ideally the gradient should be zero if there is no correlation between energy and time. The difference from zero of the gradient is divided by the uncertainty of the gradient to obtain its significance. Any significance under 3$\sigma$ is treated as statistical fluctuation, and the energy is deemed stable through the run. 

The pressure and temperature as a function of time is also measured as slow control. This is to make sure that there are no anomalies during the run, but the information is not used in the analysis. After all of the cuts are applied, the rate in the ROI is calculated. This rate should be constant as a function of time. The same gradient test as mentioned above is performed. This is done to check whether there is a significant change in rate as a function of time. The error for the rate is taken as the square root of events divided by the time. Since the rates are higher than 100 c.p.d, the Gaussian error approximation is mostly valid. 

\subsection*{Simulation}

As mentioned above a GEANT4 simulation was used to determine the geometrical efficiency for each sample. The simulation uses the known properties of the gas, the target material as well as the geometry to determine the signal in the detector. 

Before the fit, the energy spectrum is recalibrated in the following way. A probability density function (pdf) is used to fit the spectrum. The function used is the following \cite{fit_func};

\begin{align*}
  	F(E;A,\mu,\sigma,\tau) =& \left(\frac{A}{2\cdot \tau} \right) \exp\left[{\frac{(E-\mu)}{\tau}+\frac{\sigma^2}{2\cdot \tau^2}}\right] \\
	&\times \text{erfc}{\left[\frac{(E-\mu)}{\sqrt{2}\cdot \sigma}+\frac{\sigma}{\sqrt{2}\cdot\tau}\right]} \; ,
\end{align*}

where $A$ is the area of the peak, $\mu$ is the energy of the $\alpha$-particle, $\sigma$ is the Gaussian smearing of the peak and $\tau$ is the gradient of the exponentials tail. The full energy tail is described by the sum of three exponential, but since only the energy and smearing is interesting here, only a small range around the peak maximum is fitted. 

\begin{figure}[htbp]
\begin{center}
	\includegraphics[width=8.6cm]{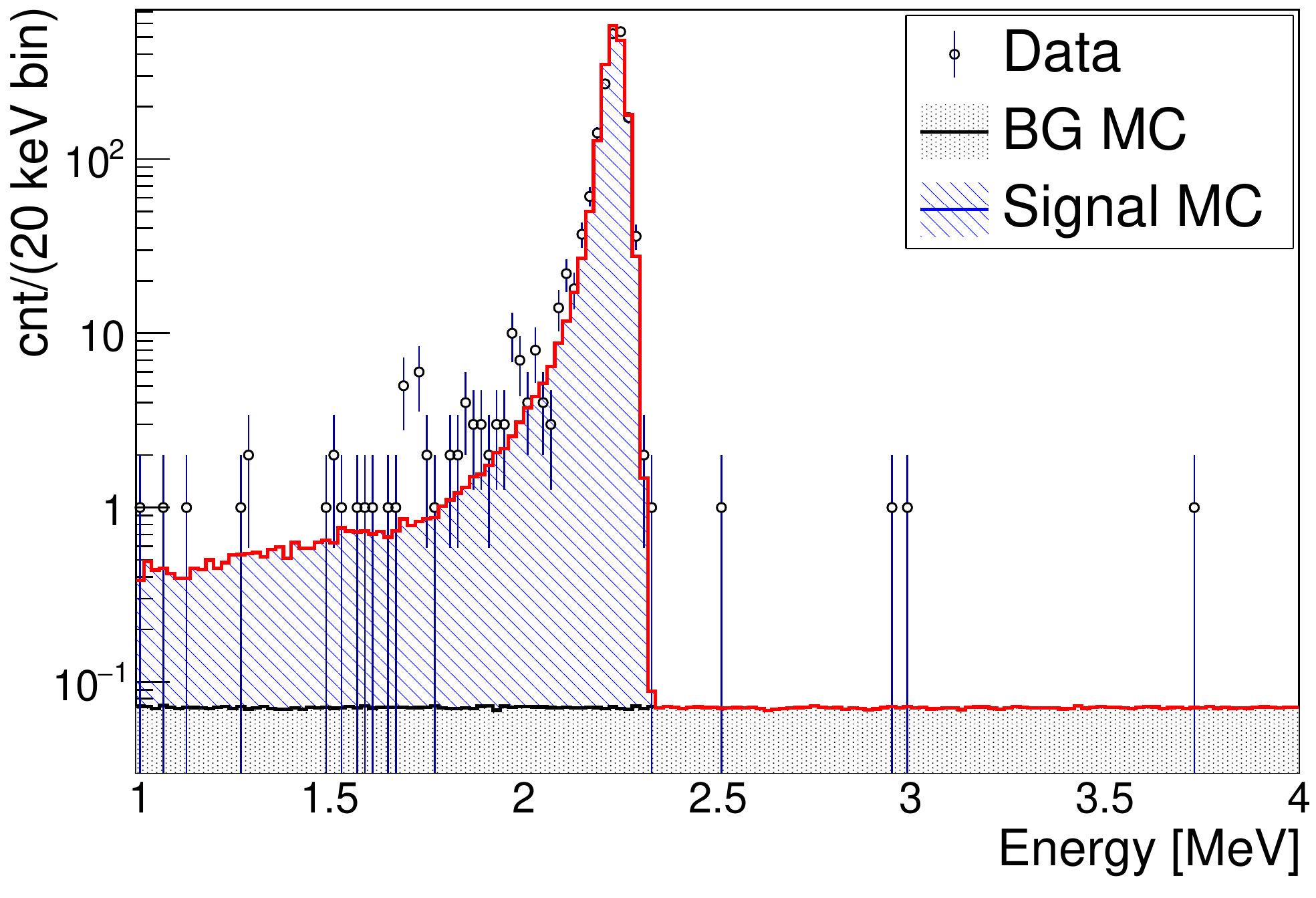}
\caption{Figure showing the energy spectrum fitted with a MC spectrum with a flat background. The fitting was done using TFractionFitter. The sample of natural samarium is 34.6(13)~nm thick. The fit has a reduced $\chi^2$ of 1.12 (165.3/148).
}
\label{fig:energy}
\end{center}
\end{figure}

The energy spectrum is then fitted with a simulated spectrum using TFractionFitter \cite{barlow} from the ROOT analysis package. A flat background spectrum is assumed as the tailing from the higher energy thorium and uranium contamination is flat at 4 MeV. This is needed to determine the fraction of signal to background events. During the fitting the algorithm varies the bin content in the simulated  spectra to take into account the statistical variation in the data. An error can occur if the algorithm is given too few simulated events to fit with. For this reason the algorithm is always fitted with at least 100 times more simulated events than the histogram being fitted. A fitted histogram is shown in Figure~\ref{fig:energy}.

The error that is given by the algorithm is usually the square root of the signal. If the error is smaller than this it is enlarged to have the minimum size of the square root of the signal. One of the problems that was encountered with the fitting simulations is that the fitter will increase the background to account for the events in the tail of the energy spectrum. This gives a smaller number of events as well as a smaller estimate for the  thickness from the energy tailing. To solve this problem, the fitting range was extended to 4~MeV. This is possible as there are very few events in the region from natural background contamination. This helps to constrain the fitter, and gives consistent fitting results. A background estimation is usually done by integrating the fitted histogram between 4~MeV and $E_{\alpha}+3\sigma$ to obtain an estimate on the number of background events expected in the ROI. This is then compared to the result from the fitter, and usually falls within one standard deviation.

\begin{figure}[htbp]
\begin{center}
	\includegraphics[width=8.6cm]{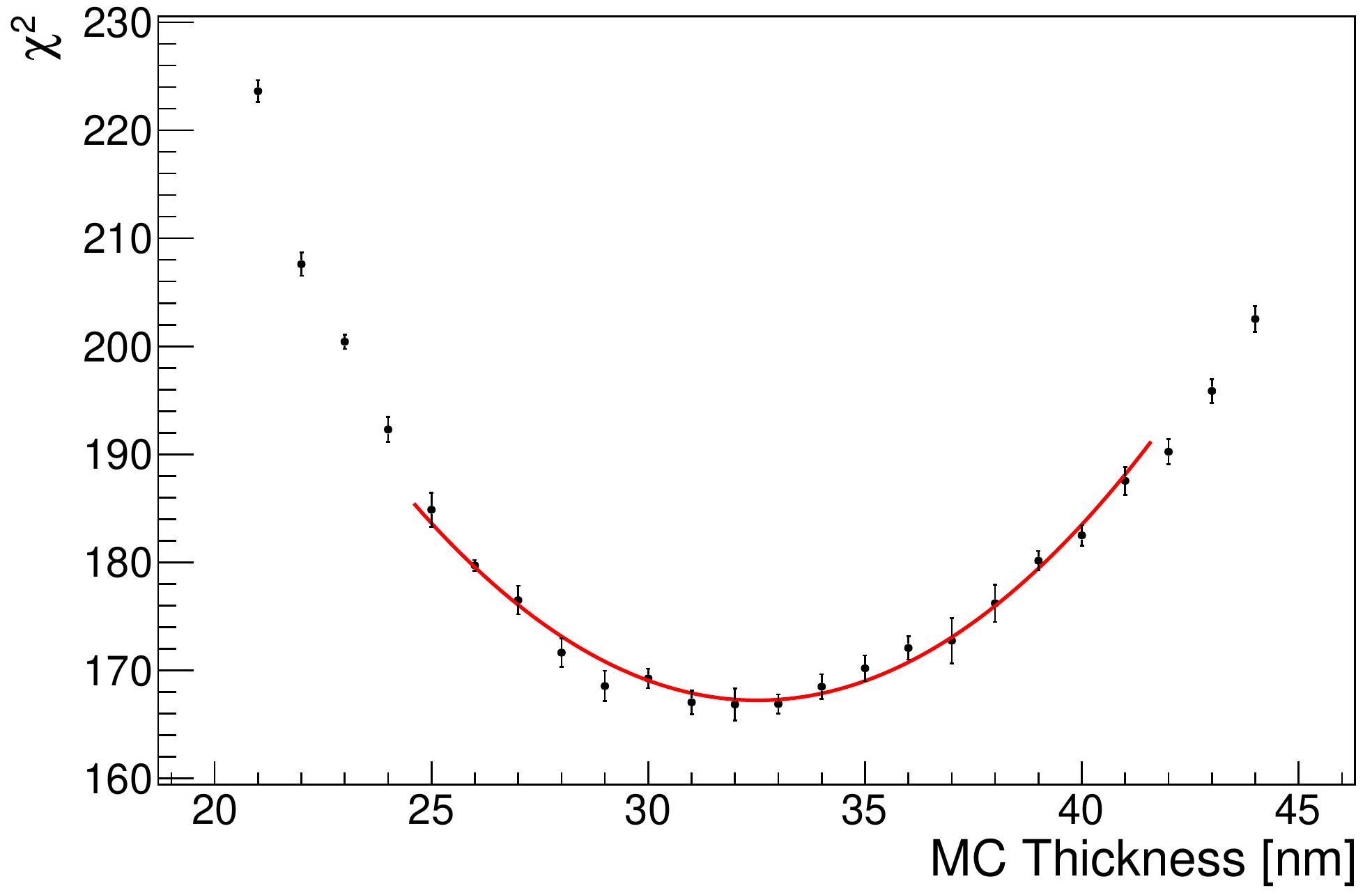}
\caption{Figure showing the $\chi^2$ of the energy spectrum fit as a function of thickness for sample SM001. To obtain the thickness and its uncertainty the $\Delta \chi^2$ = 1 and the minimum is taken from the fit (solid red line). The obtained value for the thickness from the fit is 32.5(19)~nm. This agrees with the measured thickness of 34.6(13)~nm using a density assumption of 7.54~g/cm$^3$. (The difference has a significance of less than 1~$\sigma$.) Each thickness is fitted with the simulation 20 times to determine the statistical variation, this is displayed here as the error bars.}
\label{fig:thick}
\end{center}
\end{figure}

This fitting technique is applied multiple times with a range of simulated sample thicknesses. This is done to determine the thickness that describes the data the best. This simulated thickness is important in the analysis as it is used to determine the geometric efficiency of each sample. A $\chi^2$ value is obtained with each fit over the range of simulated thicknesses. Due to the statistical nature of the smearing technique each $\chi^2$ has a variation for each thickness. Each thickness is fitted 20 times to take this variation into account. The minimum and $\Delta\chi^2 = 1$ is then obtained from the distribution. These values are taken as the simulated thickness that describes the tailing best and the 1$\cdot\sigma$ uncertainty. This approach was used because it is independent from the thickness measurement, and gives an extra check on the thickness of the samples. This is shown in Figure~\ref{fig:thick}.

To obtain the geometric efficiency from the simulated thickness is then straight forward. The simulated histogram with the best simulated thickness spectrum is integrated above 1~MeV. This is then divided by the total number of simulated events. This process is repeated for all of the simulations with thicknesses between the upper and lower error bar. The average of this is the geometric efficiency and the variation is its uncertainty.

\section{Results}

The measured half-life of the ground state decay of \iso{Sm}{147} is determined in the following manner:

\begin{equation*}
	T_{1/2} = \frac{\ln(2)}{R} \cdot N(t) \; ,
\end{equation*}

where $T_{1/2}$ is the half-life, $R$ is the true rate, corrected for the efficiency and $N(t)$ is the number of \iso{Sm}{147} atoms as a function of time. An exponential term usually describes the decay of a radio active isotope, but in this case the half-life is significantly longer than the measuring time. This means that the number of \iso{Sm}{147} atoms practically stay constant through out the measurement period. The initial number of atoms ($N_0$) is determined using the following formula;

\begin{equation*}
	N_0 = \rho_s \cdot A \cdot a\;,
\end{equation*}

where $A$ is the area of samarium deposition, $\rho_s$ is the number of samarium atom per area, and $a$ is the isotopic abundance of \iso{Sm}{147}. The true rate of the decay is measured as follows,

\begin{equation*}
	R = \frac{N_s}{T \cdot \eta} \;,
\end{equation*}

where $N_s$ is the number of signal events in the spectrum and $T$ is the runtime. $\eta$ is the probability of detecting a signal event and is the total detection efficiency (98.6(22)\%) measured with \iso{Am}{241} multiplied with the geometric efficiency obtained from the simulated spectra. The former is determined by the simulated spectrum described above and the latter is measured with a well known \iso{Am}{241} source. The measured values are displayed in Table~\ref{tab:results}.

\begin{table}[htbp]
\caption{Table of the measured rate and derived half-life for each sample using different runs. The error on the half-life is only statistical and does not include the systematic uncertainty from the detector efficiency and isotopic abundance. If a run has a delrin holder it is denoted with a ``d'', and if it had a silicon holder, it is denoted with an ``s''.The error on the events are given by the fitting algorithm. If the error is smaller than the square root of number of signal events, the error is enlarged to be the square root of number of signal events.}
\begin{tabular}{|l|l|r|c|c|c|}
\hline
Sample & Run & Time & Events &MC Eff. & T$_{1/2}$ \\
 &  & [days] &  & [\%] & [$\times{10^{11}}$ years] \\ \hline
SM001 & 1(s) & 19.1 & 1918(44) &49.288(36)  		& 1.154(34) \\   
 	    & 2(s) & 7.59 & 820(29) & 49.288(36) 		& 1.074(43) \\ \hline  
SM002 & 1(s) & 17.0 & 2011(45)   & 49.020(57) 	& 1.125(37)\\ \hline  
SM003 & 1(d) & 13.8 & 8896(102) & 47.422(43) 	& 0.988(19)\\ 
 	    & 2(d) & 5.03 & 3174(56)   & 47.422(43) 	& 1.076(17) \\ 
 	    & 3(d) & 5.80 & 4060(64)   & 47.422(43) 	& 1.096(23) \\ \hline 
SM004 & 1(d) & 1 & 2042(45) &  39.267(48)		&  1.164(27)\\  
 	    & 2(d) & 0.939 &  1892(43) & 39.267(48) 	&  1.179(29)\\ 
            & 3(s) & 14.8 &  35783(189)  &  42.749(43)	 & 1.069(10) \\ \hline 
\end{tabular}
\label{tab:results}
\end{table}

\begin{figure}[htbp]
\begin{center}
	\includegraphics[width=8.6cm]{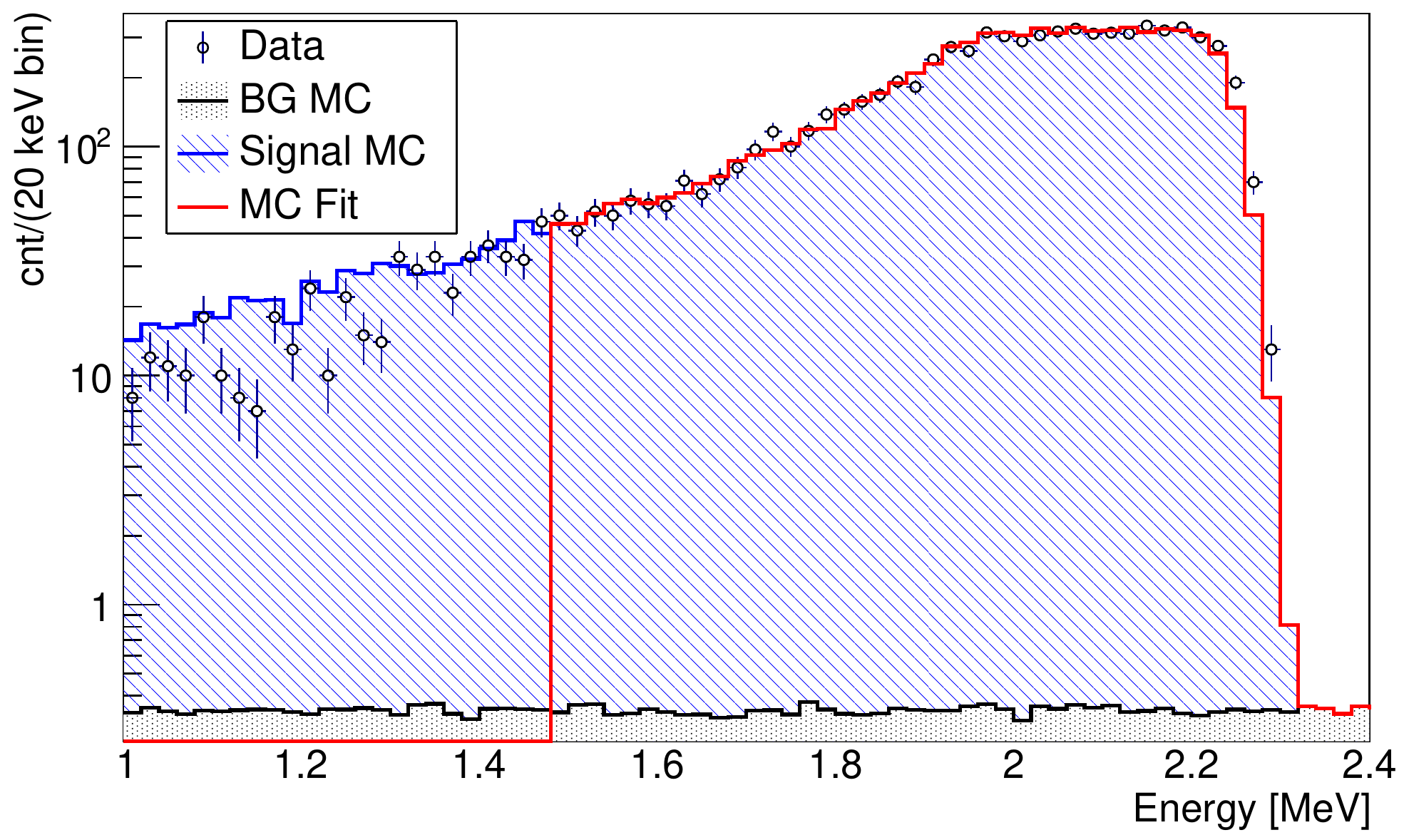}
	\includegraphics[width=8.6cm]{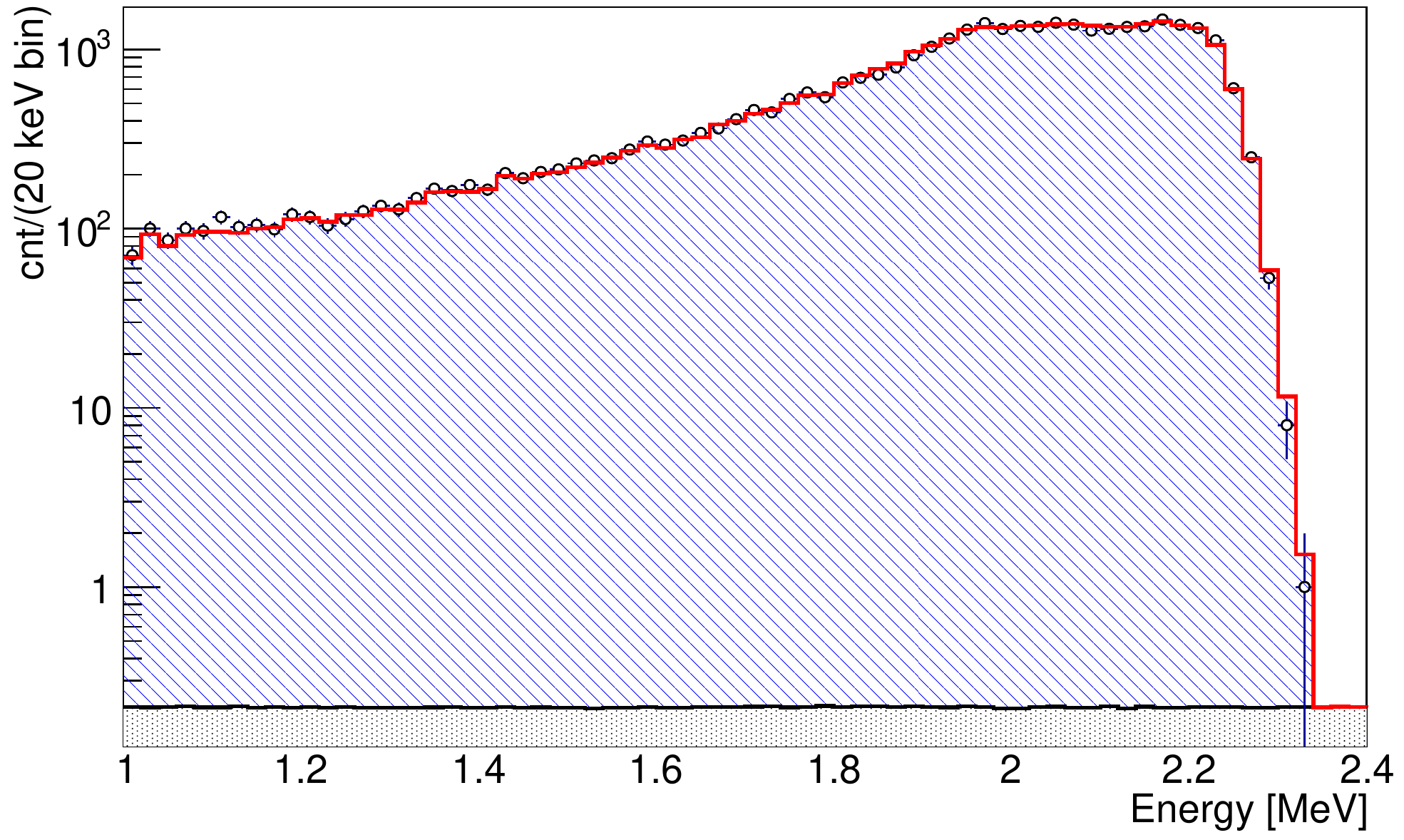}
\caption{The top figure shows the fitted energy spectrum of the entire SM004:1 run, the sample was on a delrin holder during this run. When comparing the simulation to the data, events are missing from the lower portion of the tail. The lower threshold was set to 1.48~MeV to account for this effect. The simulation thickness is given as 917(13)~nm which is in agreement with the measured value. The fit has a reduced $\chi^2$ of 1.07 (132./124). The bottom figure shows the fitted energy spectrum of run SM004:3. This run was performed on a silicon holder. It is clear from the spectrum that the simulation describes the fit well. The fit has a reduced $\chi^2$ of 0.922 (136./148) for the energy region above 1~MeV.}
\label{fig:low_thresh}
\end{center}
\end{figure}

Due to effects in the lower energy spectrum during the runs on delrin holder, the energy threshold had to be increased to 1.48~MeV. This is shown in Figure~\ref{fig:low_thresh}. The runtime of run SM004:1(d) was also shortened to 1 day due to an abnormally large change in count rate towards the end of the run. This effect is thought to be caused by the holder type. The delrin holder acts as an insulator, which means that the gas in the electric field is effectively quenched. This effect varies on a run by run basis, but the main symptom is a change in the event rate over the run. Though this is not always large enough to be detected. This phenomenon would also explain the affects seen at low energy. The events in the lower energy tail region will be emitted at a greater angle, thus are more probable to be lost in the quenched gas. It is worth noting here that it is not only the $\alpha$-decays that quench the gas, but also the cosmic-radiation.

It is clear that the delrin holder data has a significant systematic error with respect to the silicon holder data. For this reason only the runs performed on the silicon holder are used to derive the final half-life. The half-life from each measurement on a silicon holder is combined in a weighted mean. The data from the two SM001 runs were combined into a single half-life value of $1.130(31)\times{10^{11}}$~years before being used in the weighted mean. The combined half-life value is:
~ 
\begin{equation*}
		T_{1/2} = (1.0787 \pm	0.0095  (\text{stat.}) \pm	0.0244 (\text{sys.})) \times{10^{11}} \; \text{years} \;
\end{equation*}
~



The systematic uncertainty here is the combination of the uncertainty of the detector efficiency and the uncertainty on the isotopic abundance. The systematic uncertainty is dominated by the precision on the activity of the calibration source which is 2.2\%. The systematic errors are not used in the weighted mean.

\section{Decay into the first excited state of Neodymium-143}

There is a finite probability of an $\alpha$-decaying into the excited state of \iso{Nd}{143}, this leads to a de-excitation through the emission of $\gamma$-radiation. The measurement of this de-excitation would lead to a better accuracy for predicting samarium isotope $\alpha$-decay half-lives. To obtain a better sensitivity almost all of the excited state transitions have been studied using low background $\gamma$-spectroscopy. However, because of the lower Q-values and potential total angular momentum changes only the first excited state $E_{1} =$ 742.05(4)~keV \cite{gamma}, is discussed here.

Two samples of 25.05~g and 25.69~g of Sm$_2$O$_3$ were placed on a 90\% ultra-low-background high-purity germanium (ULB-HPGe) detector for 35.32 and 27.59 days respectively. The complete spectrometer including detector and shielding is described in detail in \cite{Degering}.

\begin{figure}[htbp]
\begin{center}
	\includegraphics[width=8.6cm]{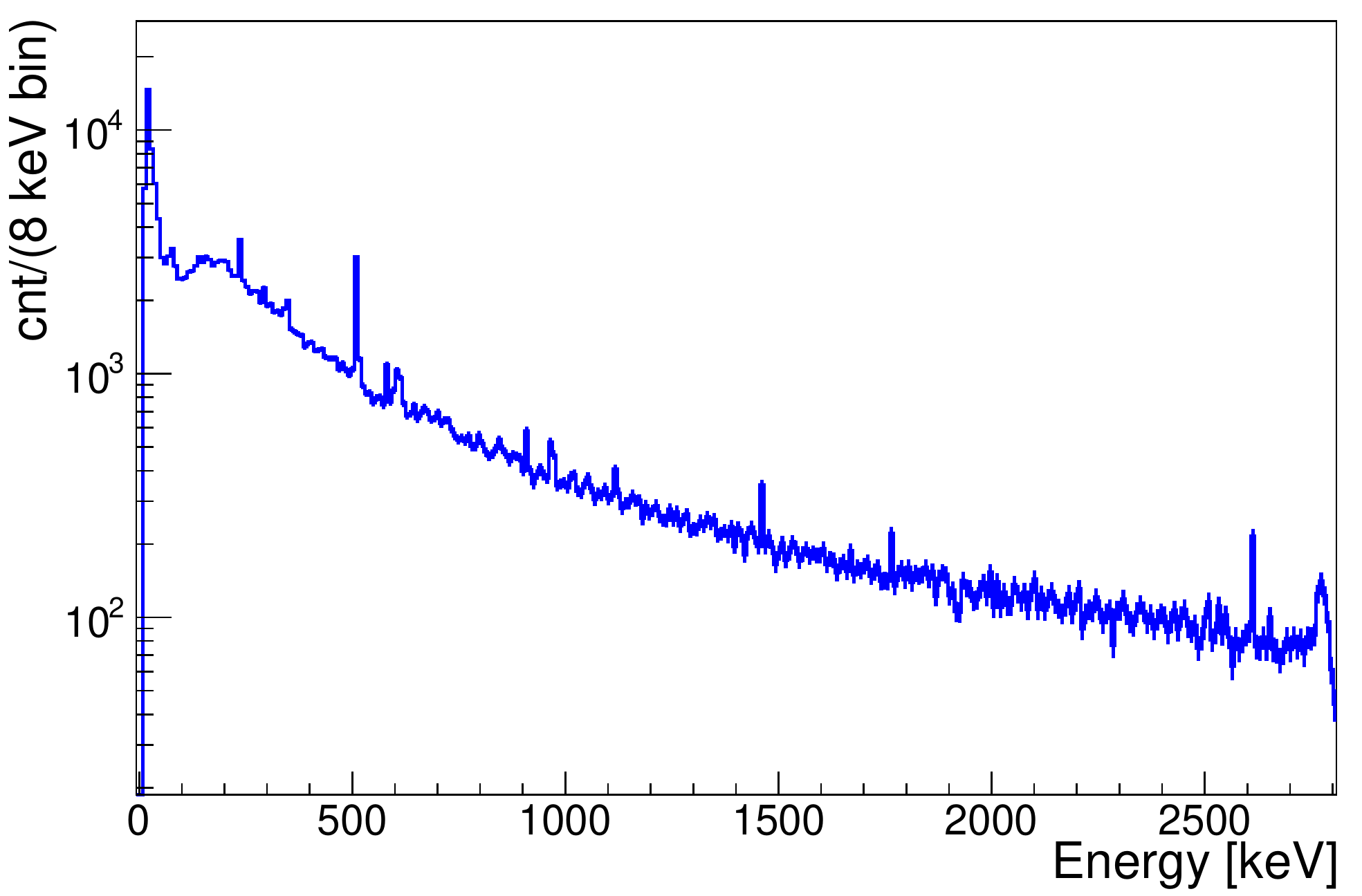}
\caption{Combined gamma spectrum of 50.74~g of Sm$_2$O$_{3}$, with a combined runtime of 62.91~days. }
\label{fig:gammafull}
\end{center}
\end{figure}

The two spectra were summed together after the calibration and were cross-checked using $\gamma$-lines from the natural radioactive contaminants. The spectrum shows a few prominent peaks. These peaks were due to contamination from natural radioactive sources like thorium and uranium decay chains as well as \iso{K}{40}. The high energy peak at 2614~keV is attributed to the decay of \iso{Tl}{208}. The positron annihilation line stemming from from muon interactions is also visible at 511~keV. This contaminations caused an overlay of Compton continua around the signal region. It was not possible to suppress the background in the measurement as it was part of the Sm$_2$O$_3$ samples. The $\gamma$-spectrum is shown in Fig.~\ref{fig:gammafull}.

\begin{figure}[htbp]
\begin{center}
	\includegraphics[width=8.6cm]{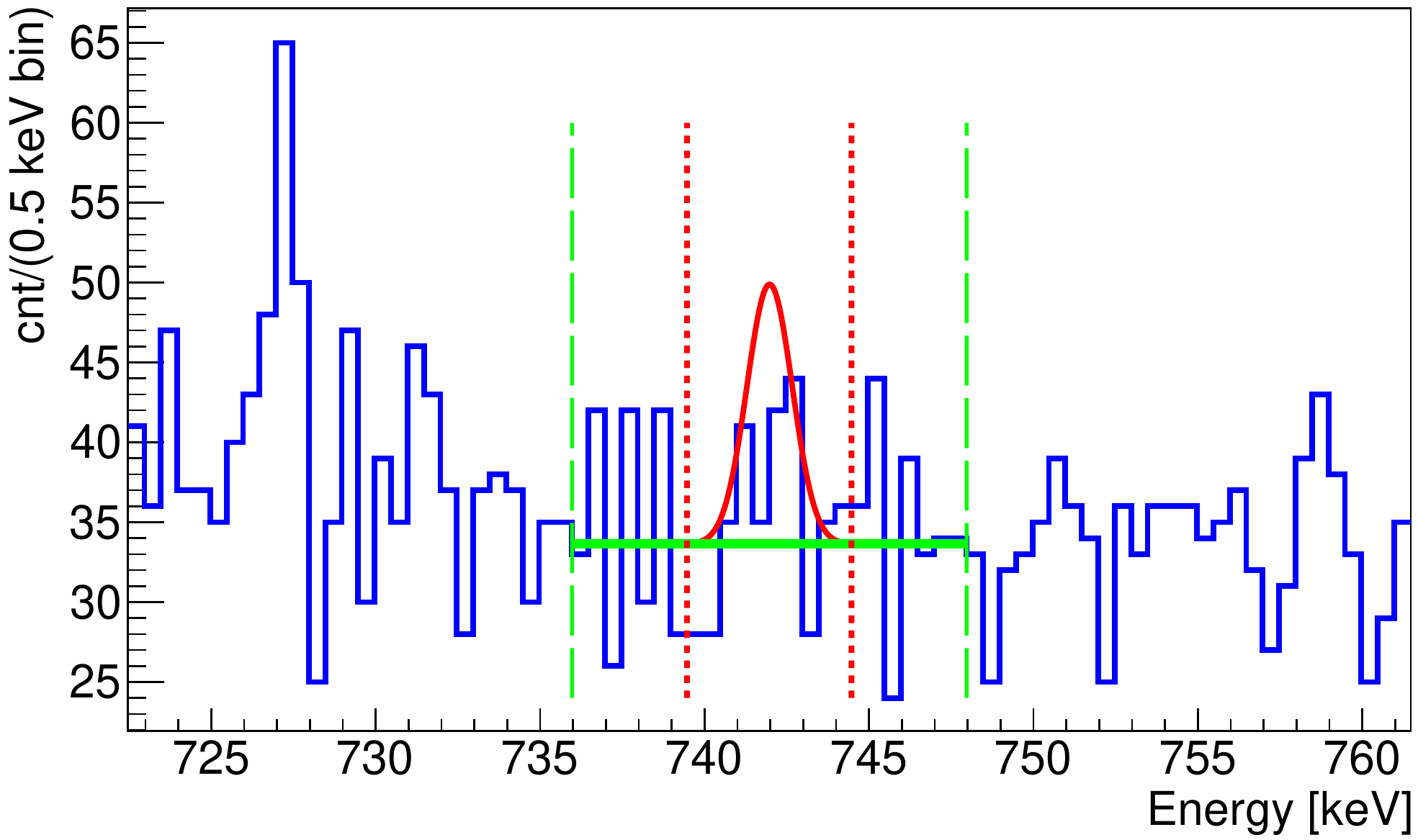}
\caption{Gamma spectrum in the region of a potential $\alpha$-decay of \iso{Sm}{147} into the first excited state of \iso{Nd}{143} at 741.98(4)~keV. The line at 727.2~keV is most likely from \iso{Bi}{212} decay from the thorium decay chain. The spectrum shows the ROI between the red finely dashed vertical lines. The background region extends to the green coarsely dashed vertical lines. The 90\% C.L. is shown on the flat background. }
\label{fig:gammapeak}
\end{center}
\end{figure}

 The peak region was split into two side bands and a region of interest (ROI) around $E_{\gamma}$, this is shown in Figure~\ref{fig:gammapeak}. The ROI is taken at $E_{\gamma}\pm 3\cdot\sigma$. The side bands were then extended for $5\cdot\sigma$ on either side to obtain a background estimate in the ROI. At $E_{\gamma}$ the energy resolution was determined to be 0.7~keV. A simulation was made with MAGE\cite{mage} which uses the GEANT4 simulation package framework to obtain the full energy peak efficiency. The simulation showed the efficiency at 742 keV is 5.5\%. The side bands were slightly extended to fit the binning of the histogram. The side bands were fitted with a first order polynomial, the gradient of the polynomial showed a statistical significance of 0.0093~$\sigma$ and thus the background was treated as a uniform distribution. The fit gave a background estimate in the ROI of 336(11)~counts. The measured number in the ROI was 352~counts.

To obtain an upper limit on the number of counts the TRolke package\cite{rolke} from ROOT was used. A Gaussian uncertainty in the background estimate with a known efficiency was assumed. The upper limit in the signal was determined to be 53 events at 90\% C.L. in the ROI. This gives a lower half-life limit on the decay of \iso{Sm}{147} to the excited state of \iso{Nd}{143} of $3.3\times10^{18}$~years. This result assumes that the samarium has natural abundance taken from \cite{deLaeter2003}. 

The theoretical prediction for this decay mode is taken from the Viola-Seaborg relation between the Q-value and half-life\cite{Viola}. This is done here by subtracting the energy of the excited state from the Q-value of the transition. This gives an expected value for the half-life of the order of $10^{24}$ years. This is a rough estimate as the change in total angular momentum is not taken into account, but would make the expected half-live even longer. The small Q-value is expected to have a much greater impact on the half-life. To reach the predicted half-life region lower background and larger exposure (mass, time) is needed. 

\section*{Conclusion}

The half-life of \iso{Sm}{147} was determined using an ultra low background TF-GIC. Even though the activity of the samples are very small, the rates are still far above the background rate. Special attention has been given to the tailing caused by the self absorption. A Monte Carlo simulation was developed to describe the energy tailing. The simulation did not only describe the tailing perfectly through the whole range of thicknesses, but was also used to validate the thickness of the samples. 

A great deal of work has gone into the characterisation of the samples, using techniques that are well established and understood. All of the measurement methods were complimentary and validated each other. The isotopic abundance of the material was measured using ICP-MS. The abundance was found to be in good agreement with literature value. 

The counting efficiency of the TF-GIC was measured in various ways. A detailed paper was published on this subject \cite{cham}. The energy spectra were analysed using the ROOT class TFractionFitter. This allowed for the proper extraction of background rates in the fitting as well as the correct geometrical efficiency from the simulation. 

\begin{figure}[htbp]
\begin{center}
	\includegraphics[width=8.6cm]{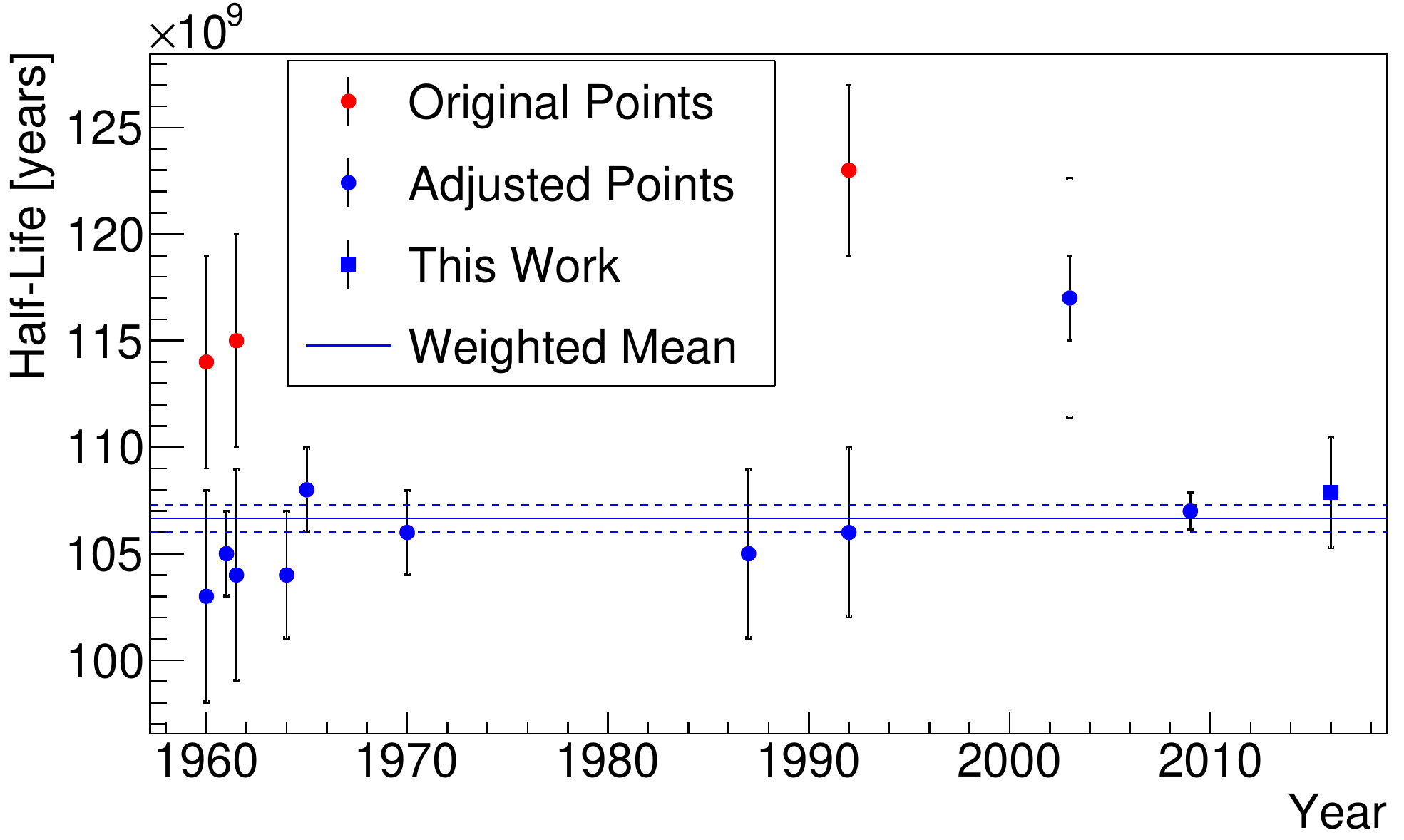}
\caption{Half-life values plotted as a function of year. The values are taken from \cite{A147}. The red circular points mark the original published results. The blue circular points mark the adjusted half-life values, this is made by following the method described in \cite{A147}. The weighted mean is calculated using the method described in \cite{raj} with the adjusted half-life values. The errors used for this weighed mean are shown in square braces, the used values are shown in Table~\ref{tab:hl}. }
\label{fig:halflives}
\end{center}
\end{figure}

Figure~\ref{fig:halflives} shows a graph of \iso{Sm}{147} measurements as a function of year as well as the value obtained in this work. Our value of 1.079(26)$\times{10^{11}}$~years is in very good agreement ( $< 1 \sigma$) with the value obtained by Kossert et al. (2009) \cite{kos09} and is at slight tension with the value obtained by Kinoshita et al. (2003) \cite{kin}. Possible sources for the discrepancy of this measurement have already been discussed in \cite{kos09}. 

An evaluation of the half-lives of \iso{Sm}{147} was performed using the Rajeval technique \cite{raj} on the values given in \cite{A147} as well as the addition of the measurement from Kossert et al.. Three of the values have been adjusted according to the correction factors mentioned in \cite{A147} and \cite{kos09} to account for the contribution from the decay of \iso{Sm}{146}. The Rajeval technique increases the uncertainty of the data points that are not consistent with the group mean. In this case the uncertainty of Kinoshita et al. was increased by a factor of 2.8. The weighted mean from this was found to be $1.067(6)\times{10^{11}}$~years. Without the adjustment of the half-life mentioned before the $\chi^2$/(n-1) of the group would be 2.60, after the adjustment the $\chi^2$/(n-1) is 0.65 which indicated a better agreement between half-life values. All of the values used to derive the weighted mean are given in Table~\ref{tab:hl}. 
\\
\begin{table}[h!]
\caption{Table showing the historic half-life measurements of $^{147}$Sm. All of the half-lives are taken from \cite{A147}, and the augmentations are made using the recommendations also mentioned in \cite{kos09}.}
\begin{tabular}{| r r | c | c |}
\hline
Reference& &     Value          & Augmented  \\ 
		&&    [$\times{10^{11}}$~a]  & Value [$\times{10^{11}}$~a] \\  \hline
(2009Ko01)& \cite{kos09}			& 1.070(9) 	&$\cdot$ 			\\ \hline
(2003Ki26) &\cite{kin} 			& 1.17(2) 		&  1.17(6) 			\\ \hline
(1992Ma26)& \cite{Martins1992} 	& 1.23(4)		& 1.06(4)   \cite{Begemann} \\ \hline
(1987Al28)& \cite{alb} 			& 1.05(4)  		&$\cdot$ 			\\ \hline
(1970Gu14)&\cite{GUPTA19703425} & 1.06(2) 		&$\cdot$ 			\\ \hline
(1965Va16)&\cite{1965Va16} 		& 1.08(2)  		&$\cdot$ 			\\ \hline
(1964Do01)&\cite{Donhoffer1964489} & 1.04(3)  	&$\cdot$ 			\\ \hline
(1961Ma05)&\cite{Mac} 			& 1.15(5) 		& 1.04(5) 			\\ \hline
(1961Wr02)& \cite{Wright}			& 1.05(2)  		& $\cdot$ 			\\ \hline
(1960Ka23)& \cite{Karras} 		& 1.14(5)  		&1.03(5) 			\\ \hline
\end{tabular}
\label{tab:hl}
\end{table}

\section*{Acknowledgments}

The authors would like to thank Mih\'aly Braun for his ICP-MS measurement of the isotopic abundance of the samarium. The authors would also like to thank Bj\"{o}rn Lehnert for his help in developing a MAGE simulation to determine the full energy peak efficiencies for the $\gamma$-decay. We also thank Katja Berger for measuring the sample area. Furthermore, the authors thank Ren\'e H\"ubner for measurement of the TEM images. We would like to thank the referee for their helpful suggestions in improving the measurement in this work. Parts of this research were carried out at the Ion Beam Center at the Helmholtz-Zentrum Dresden - Rossendorf e. V., a member of the Helmholtz Association. This work is supported by BMBF under contract 02NUK13A HZDR and 02NUK13B TUD.

\bibliography{samarium.bib}

\end{document}